\documentclass[11pt]{article}
\usepackage{xcolor}
\definecolor{red}{rgb}{0,0,0} 

\usepackage{fullpage}
\usepackage{setspace}
\usepackage{parskip}
\usepackage{titlesec}
\usepackage[section]{placeins}
\usepackage{xcolor}
\usepackage{breakcites}
\usepackage{lineno}
\usepackage{times}

\usepackage{apacite}

\PassOptionsToPackage{hyphens}{url}
\usepackage[colorlinks = true,
            linkcolor = blue,
            urlcolor  = blue,
            citecolor = blue,
            anchorcolor = blue]{hyperref}
\usepackage{etoolbox}
\makeatletter
\patchcmd\@combinedblfloats{\box\@outputbox}{\unvbox\@outputbox}{}{%
  \errmessage{\noexpand\@combinedblfloats could not be patched}%
}%
\makeatother

\usepackage{natbib}

\renewenvironment{abstract}
  {{\bfseries\noindent{\abstractname}\par\nobreak}\footnotesize}
  {\bigskip}

\titlespacing{\section}{0pt}{*3}{*1}
\titlespacing{\subsection}{0pt}{*2}{*0.5}
\titlespacing{\subsubsection}{0pt}{*1.5}{0pt}

\usepackage{authblk}

\usepackage{graphicx}
\graphicspath{{figures/}}
\usepackage[space]{grffile}
\usepackage{latexsym}
\usepackage{textcomp}
\usepackage{longtable}
\usepackage{tabulary}
\usepackage{booktabs,array,multirow}
\usepackage{amsfonts,amsmath,amssymb}

\newif\iflatexml\latexmlfalse

\AtBeginDocument{\DeclareGraphicsExtensions{.pdf,.PDF,.eps,.EPS,.png,.PNG,.tif,.TIF,.jpg,.JPG,.jpeg,.JPEG}}

\usepackage[utf8]{inputenc}
\usepackage[english]{babel}
\usepackage{float}
\usepackage[margin=1.5in]{geometry}

\newcommand{\btheta}{\mbox{\boldmath $\theta$}}
\renewcommand{\d}{{\bf d}}
\newcommand{\n}{{\bf n}}
\newcommand{\h}{{\bf h}}

\newcommand{\e}{{\rm e}}
\renewcommand{\S}{{\bf S}}

\begin{document}

\title{Computational Techniques for  Parameter Estimation of  Gravitational Wave Signals }

\author[1]{Renate Meyer}%
\affil[1]{Department of Statistics, The University of Auckland, Auckland, New Zealand}
\author[1]{Matthew C.~Edwards}%
\author[1,2]{Patricio Maturana-Russel}%
\affil[2]{Department of Mathematical Sciences, Auckland University of Technology, Auckland, New Zealand}%
\author[3]{Nelson Christensen}%
\affil[3]{Artemis, Observatoire de la C\^{o}te d'Azur, Universit\'{e} C\^{o}te d'Azur, Nice, France}
\vspace{-1em}

  \date{}

\begingroup
\let\center\flushleft
\let\endcenter\endflushleft
\maketitle
\endgroup

\selectlanguage{english}
\begin{abstract}
Since the very first detection of gravitational waves from the coalescence of two black holes in 2015, Bayesian statistical methods have been routinely applied by LIGO and Virgo to extract the signal out of noisy interferometric measurements, obtain point estimates of the physical parameters responsible for producing the signal, and rigorously quantify their uncertainties. Different computational techniques have been devised  depending on the source of the gravitational radiation and the  gravitational waveform model used. Prominent sources of gravitational waves are binary black hole or neutron star mergers, the only objects that have been observed by detectors to date. But also gravitational waves from core collapse supernovae, rapidly rotating neutron stars, and the stochastic gravitational wave background are in the sensitivity band of the ground-based interferometers and expected to be observable in future observation runs. As nonlinearities of the complex waveforms and the high-dimensional parameter spaces preclude analytic evaluation of the posterior distribution, posterior inference for all these sources relies on computer-intensive simulation techniques such as Markov chain Monte Carlo methods. A review of state-of-the-art Bayesian statistical parameter estimation methods will be given for researchers in this cross-disciplinary area of gravitational wave data analysis.

\end{abstract}%

\par\null

This is the peer reviewed version of the following article: ``Computational Techniques for Parameter Estimation of Gravitational Wave Signals", which has been published in final form at \href{URL}{http://dx.doi.org/10.1002/wics.1532}. This article may be used for non-commercial purposes in accordance with Wiley Terms and Conditions for Use of Self-Archived Versions.\pagebreak

\section{Introduction}

{\label{intro}}





The era of observational gravitational wave astronomy truly began with the detection of GW150914 --  gravitational waves from the inspiral and merger of two stellar-mass black holes that coalesced to form a single rotating black hole --  by the two Advanced Laser Interferometer Gravitational-wave Observatory (Advanced LIGO) detectors \citep{TheLIGOScientific:2016qqj} in 2015.
Even though the existence of gravitational waves had already been predicted by Einstein in 1916 \citep{Einstein:1916a} as a consequence of the theory of General Relativity, only indirect evidence had so far been provided by 
radio observations of the binary pulsar PSR1913+16 in 1974 and the double pulsar PSR J0737-3039 in 2003. \cite{1982ApJ...253..908T} showed that the energy loss associated with the orbital decay rate was consistent with the emission of gravitational waves. The discoverers of PSR1913+16 \citep{1975ApJ...195L..51H}, Joseph Taylor and Russell Hulse, were awarded the Nobel Prize in Physics in 1993. The groundbreaking first  direct measurement of gravitational waves earned the founders of LIGO, Rainer Weiss, Barry Barish, and Kip Thorne, the Nobel Prize in Physics in 2017.  Since then, many more black hole mergers have been observed in subsequent observation runs of the Advanced LIGO and Virgo interferometers \citep{LIGOScientific:2018mvr,AbbottR2020GGWf,GW190412,Abbott:2020uma}. The detection of the neutron star merger GW170817 \citep{TheLIGOScientific:2017qsa}, seen both in gravitational and electromagnetic waves, ushered in the new era of gravitational wave multi-messenger astronomy \citep{2017ApJ...848L..12A}.

Gravitational waves are produced by non-axisymmetric acceleration of mass, such as two compact neutron stars orbiting each other.
They are quadrupole (lowest order) waves that propagate outwards from their source at the speed of light. Unlike electromagnetic waves, they pass through any matter. Their effect is orthogonal to the direction of propagation. They have two polarizations, a plus- and cross- polarization. The plus-polarization stretches the distance of two points on the horizontal axis and simultaneously compresses the distance between two points on the vertical axis.  The cross-polarization has a similar effect, but rotated by 45 degrees. 
Gravitational wave detectors, such as Advanced LIGO \citep{TheLIGOScientific:2014jea} and Advanced Virgo \citep{TheVirgo:2014hva}, are based on Michelson laser interferometry \citep{HariharanP.2007BoI,MatthewPitkin2011GWDb}. 
A beam of light is split in two and sent in two equal-length (4 km for LIGO, 3 km for Virgo) perpendicular arms (in vacuum) with mirror-coated test masses suspended on pendulums at each end of the arms, storing the light and increasing the effective arm length by a factor of $\sim 300$ as the light bounces back and forth. The light beams exiting the arms are recombined. A passing gravitational wave changes the lengths of each arm and thus the interference pattern measured by photodetectors.
The wave amplitude,
the dimensionless {\em strain}, denoted by $h$, is measured by the relative change in spacing $\frac{\Delta L}{L}$ between two test masses where $L$ denotes the equilibrium spacing.
The strain is proportional to the second time derivative of the source quadrupole moment
and decreases in proportion to the inverse distance from the source \citep{SchutzBernardF.2009Afci}.
 Thus, in practice only gravitational waves from massive and rapidly accelerating objects in the Universe will be detectable. Gravitational waves from astronomical sources,  due to their large distance from  Earth, have detectable strains of the order of  $10^{-21}$.

Here, we focus on observations from the network of ground-based interferometers.  After a five-year long upgrade, Advanced LIGO  became operational in 2015 with two detectors, in Hanford, Washington and Livingston, Louisiana, and Advanced Virgo with a detector in Cascina, Italy, in 2017. A third LIGO detector is planned to be built in India \citep{Unnikrishnan:2013qwa}, and the Japanese underground cryogenic detector KAGRA \citep{SomiyaKentaro2012Dcok} is currently coming online with its commissioning activities.
With a single detector, it is difficult to 
determine the
sky location of a transient source. A world-wide network of interferometers is important to estimate the source position using timing, amplitude, and polarization information of the signal \citep{BizouardMarie2013Sfgw}. The Einstein Telescope \citep{SathyaprakashB2012SooE, MaggioreMichele2020Scft} is a proposed third generation underground cryogenic detector with 10km 
arm lengths in a triangular formation. \textcolor{red}{The Cosmic Explorer, a planned 40 km  L-shaped detector, will greatly enhance the sensitivity due to the significantly increased arm lengths \citep{AbbottBP2017Etso}.}  Pulsar timing arrays~\citep{DahalPravinKumar2020Ropt} are sensitive to low-frequency gravitational waves in the range of 
$10^{-9}$ to $10^{-6}$ Hz. Furthermore, a space-based interferometer, the so-called Laser Interferometer Space Antenna (LISA)  is planned to be launched in 2034 by the
European Space Agency with three satellites 2.5 million km apart, forming an equilateral triangle in an Earth-trailing, heliocentric orbit \citep{2017arXiv170200786A}. A space-based interferometer would be sensitive to frequencies from 0.1 mHz to 1 Hz. Because LIGO and Virgo are currently operating, and making detections, this review is concentrating on parameter estimation methods for these ground based detectors. However the parameter estimation development for LISA has been long and active~\citep{PhysRevD.72.022001,Umst_tter_2005,Cornish:2005qw,Crowder:2006eu,Babak:2007zd,Stroeer:2007tg,Arnaud:2007vr,Ali:2012zz,Baghi:2019eqo,KatzMichael2020Gmbh,ToubianaAlexandre2020Peos}. LISA will actually observe thousands of simultaneous signals, and therefore pose one of the biggest parameter estimation challenge ever in physics~\citep{2017JPhCS.840a2026B,Robson_2017,Cornish_2017,Littenberg:2020bxy,MarsatSylvain2020EtBp,CornishNeil2020BHHw}.

Even though much information about the emitting source of the gravitational waves can be learned from the direct inspection of the observed gravitational waveforms  \citep{NoAuthor2017}, the full posterior distribution of the source parameters using Bayesian computational techniques is required to accurately estimate the properties of the source, such as the masses and spins of the two black holes, and to quantify associated uncertainties. A review of state-of-the-art parameter estimation methods will be the focus of this paper. As this is not only an expansive but also a rapidly evolving research area, we can only strive for a comprehensive review but cannot claim exhaustiveness. For details regarding the computational methodology and their implementations, we refer to the relevant statistical and machine learning literature.

\bigskip

\section{Parameter Estimation}
\label{sec:PEgeneral}

Following a calibration procedure for each detector, the time series of dimensionless strain measurements are produced.
The strain is the fractional change in spacings between two test masses due to a passing gravitational wave, that is the difference in lengths relative to the total arm length of the interferometer.
The \textcolor{red}{calibrated} observations $\d^{(k)}=(d^{(k)}(t)), t=1,\ldots,T$ of gravitational waves from detector $k$ are modelled as a deterministic signal, the strain $\h^{(k)}=\left( h^{(k)}(t|\btheta)\right), t=1,\ldots,T$ plus additive interferometer noise 
$n^{(k)}(t)$, that is,
\begin{equation}\label{modelintime}
d^{(k)}(t) = h^{(k)}(t|\btheta) + n^{(k)}(t),\quad t=1,\ldots,T.
\end{equation}
Thus, the model consists of two parts: a model for the gravitational wave signal and a model for the noise, both equally important for parameter estimation \citep{LIGOScientific:2019hgc}.

The interferometers are subject to a variety of {\em noise} components including quantum noise, seismic noise, thermal noise, and gravity gradient noise. Also, environmental noise or malfunctioning of equipment can cause transient noise events \citep{Abbott_2016,NuttallL2018Ctni}, or noise spectral lines \citep{PhysRevD.97.082002}. All these combined are modelled by the noise time series $n^{(k)}(t)$,
usually assumed to be a Gaussian, stationary time series with zero mean  and covariance matrix $\Sigma_k$.

The assumptions on the  noise  determine the form of the likelihood.
The observations  from different detectors in a network of $K$ detectors are usually assumed to be independent and thus, for a coherent analysis that includes data from all detectors, the joint likelihood is the product of the individual likelihood functions:
\begin{equation}\label{jointlikelihood}
L(\d|\btheta)= \prod_{k=1}^K L^{(k)}(\d^{(k)}|\btheta)=  \prod_{k=1}^K \frac{1}{\det(2\pi\Sigma_k)^{1/2}}\e^{-\frac{1}{2}(\d^{(k)}-\h^{(k)})^{\top}\Sigma_k^{-1}(\d^{(k)}-\h^{(k)})  }
\end{equation}

Usually, parameter estimation is implemented in the frequency domain because after a Fourier transform, instead of a  multivariate Gaussian likelihood with non-diagonal covariance matrix, the complex vector $\tilde{\d}^{(k)}$ containing Fourier coefficients defined by
\[\tilde{d}^{(k)}_j=\tilde{d}^{(k)}(f_j)=\sum_{t=1}^{T} d^{(k)}(t) \e^{-itf_j}\] 
at the  Fourier frequencies $f_j=2\pi j/T, j=0,\ldots,N$, $N= \lfloor(T-1)/2 \rfloor$,
is approximately (for large $T$) a complex multivariate Gaussian but with a {\em diagonal} covariance matrix $S^{(k)}$
that contains the power spectral density $S^{(k)}(f_j)$  at the Fourier frequencies $f_j$ on the diagonal.
 The power spectral density 
\begin{equation}\label{eq:S-gamma}
S^{(k)}(f)=\sum_{\ell=-\infty}^{\infty} \gamma^{(k)}(\ell) \e^{-i\ell f}
\end{equation}
is the
Fourier transform of the autocovariance function $\gamma^{(k)}(l)$ of the stationary \textcolor{red}{noise} time series. The diagonal covariance structure in this  so-called {\em Whittle} likelihood approximation \citep{WhittleP1957CaPS, LeeSamuelFinn1992DMaG, Cutler1994Gwfm,  KirchClaudia2019BWNC}
 facilitates the calculation of the inverse and determinant  in the Whittle likelihood:

\begin{equation}\label{Whittlelikelihood}
L(\d|\btheta)\approx 
 \prod_{k=1}^K \frac{1}{\det( \pi T\S^{(k)})}\e^{-\frac{1}{T}(\tilde{\d}^{(k)}-\tilde{\h}^{(k)})^*\S^{(k)^{-1}}(\tilde{\d}^{(k)}-\tilde{\h}^{(k)})} 
\end{equation}

\begin{figure}
\centering
\includegraphics[width=0.9\textwidth]{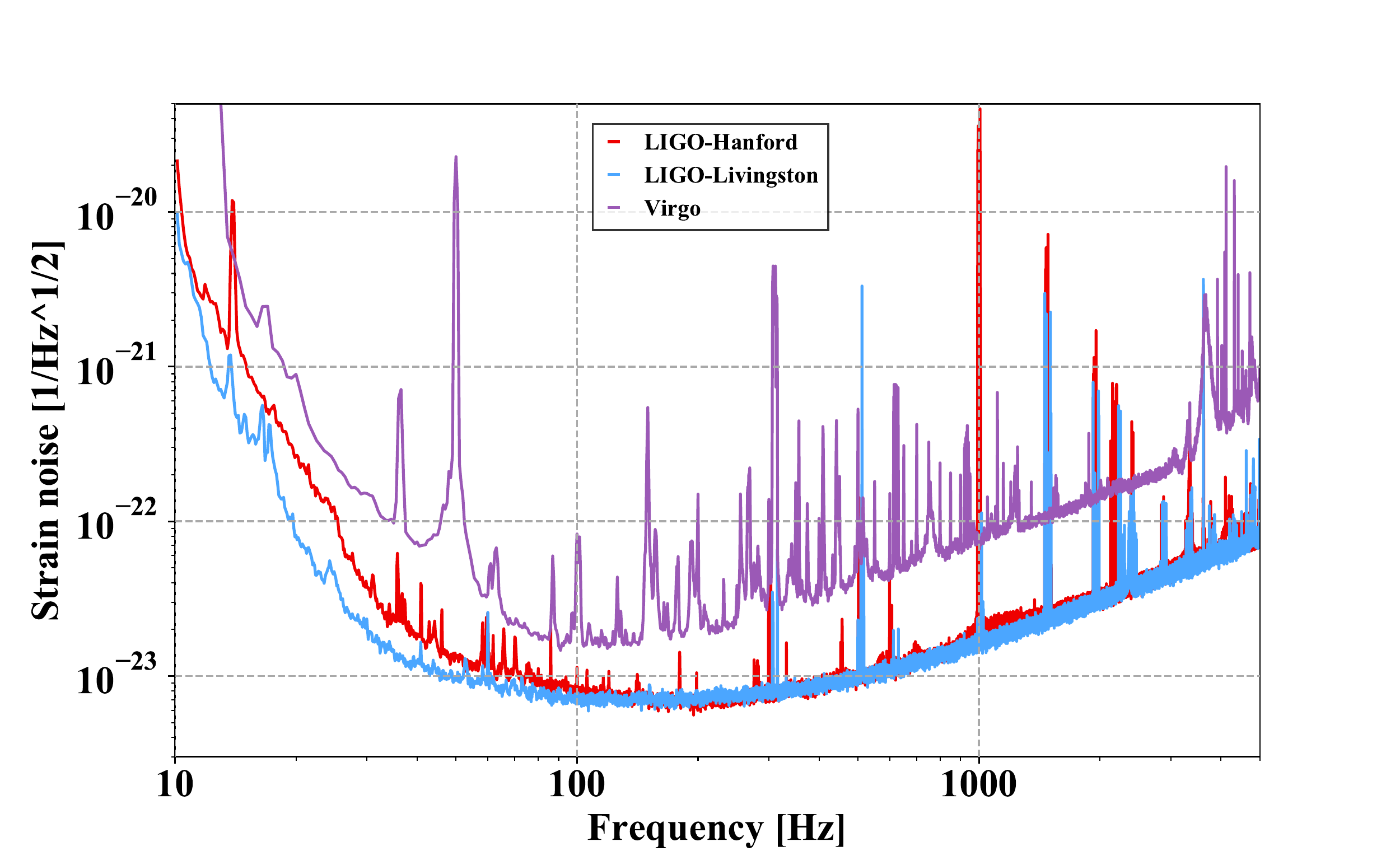}
\caption{The Advanced LIGO and Advanced Virgo sensitivity curves. These are typical of the best sensitivities of the detectors during their second observational run~\citep{LIGOScientific:2018mvr}. Source: LIGO Scientific Collaboration}
\label{fig:sensitivity}
\end{figure}

In the following, we assume that the power spectral density (PSD) for each detector is known  but in Section \ref{sec:noise} we outline procedures to deal with an unknown noise spectral density that is estimated simultaneously with the signal parameters. The stationarity assumption is usually adequate for \textcolor{red}{transient}  signals such as 
the signal from a merger of two black holes but it is well known that the interferometer noise is slowly time-varying over longer periods of about one minute \citep{ChatziioannouKaterina2019Nsem}.
Therefore, it will be important to take the time-varying noise into account when estimating the parameters of gravitational wave signals of longer duration such as those produced by neutron star mergers or pulsars.
Furthermore, the power spectrum contains many high power narrow spectral lines which are not compatible with the Gaussian assumption. Many of these are due to known sources such as the 60~Hz power line harmonics, so-called `violin modes' caused by thermally excited mirror suspension and their harmonics, or calibration lines inserted by moving the end mirrors~\citep{PhysRevD.97.082002}. \textcolor{red}{\citet{BerryChristopherP.L.2015Pefb} perform a systematic study on the performance of parameter estimates under the Gaussian assumption but with real, non-ideal noise.}
Figure \ref{fig:sensitivity} shows the amplitude spectral density (the square root of the power spectral density) of both Advanced LIGO detectors and the Advanced Virgo detector from their second observational run \citep{LIGOScientific:2018mvr}. For the purpose of parameter estimation,
the power spectral density is usually \textcolor{red}{assumed to be known and fixed. Its value is usually obtained  by one of two methods.} The Welch method \citep{WelchP1967Tuof} uses a  separate stretch of data close to  but not containing the 
time period of the signal. This stretch of data is divided into overlapping segments of the same duration $T$ as the signal period.
These segments are Fourier transformed using the FFT algorithm after windowing with a Tukey or Hanning window to avoid spectral leakage~\citep{LIGOScientific:2019hgc}. The Welch method 
 averages over the periodograms of each individual  time segment to reduce the variance of the estimate. \textcolor{red}{Often, the median instead of mean of the periodograms is used as it is more robust with respect to outliers \citep{Veitch:2014wba}. As an alternative to this off-source Welch method, an on-source estimate, based on the same data that contains the signal, is obtained by BayesWave \citep{Cornish:2014kda}, described in Section \ref{sec:noise}.}
The power spectral density is then assumed to be known and equal to the Welch or BayesWave estimate when estimating the parameters of the signal.  The likelihood Eq. (\ref{Whittlelikelihood}) for \textcolor{red}{known} PSD simplifies to
\begin{equation}
\label{WhittlelikelihoodfixedPSD}
L(\d|\btheta)\propto 
 \prod_{k=1}^K \e^{-\frac{1}{T}(\tilde{\d}^{(k)}-\tilde{\h}^{(k)})^*\S^{(k)-1}(\tilde{\d}^{(k)}-\tilde{\h}^{(k)})} 
\end{equation}
 \textcolor{red}{Alternative parametric and nonparametric Bayesian estimates of the power spectral density are discussed in Section \ref{sec:noise}.}

The {\em strain signal} is characterized by various parameters (depending on the gravitational wave source) such as the masses and spins of
the progenitors in case of a compact binary coalescence, compiled in the parameter vector $\btheta=(\theta_1,\ldots,\theta_p)$.
Considering a geocentric reference frame, the strain measured at detector $k$ of a gravitational wave source with polarization amplitudes $h_+$ and $h_\times$ located in the sky at 
$(\alpha,\delta)$ where $\alpha$ is the right ascension and $\delta$ the declination of the source is 
\begin{equation}\label{signalintime}
h^{(k)}(t|\btheta)=F^{(k)}_+(\alpha,\delta,\psi)h_+(t-\tau^{(k)}|\btheta) +F^{(k)}_\times(\alpha,\delta,\psi)h_\times(t-\tau^{(k)}|\btheta).
\end{equation}
 $F_{+,\times}$ are the antenna response functions that depend on the source locations, the polarization angle $\psi$  of the waves \citep{PhysRevD.46.5250,PhysRevD.63.042003}, plus the locations and orientations of the detectors.
For short transient signals, the time dependence of the antenna response functions due to the rotation of the earth can be safely ignored and assumed to be constant throughout the observation period, but needs to be \textcolor{red}{taken into account} for long signals. 
The parameter $\tau^{(k)}=\tau^{(k)}(\alpha,\delta)$ denotes the location-dependent time delay.
\textcolor{red}{
	For a detailed explanation of the calibration of the gravitational wave detectors and methods to take the associated calibration uncertainty into account, we refer to \citet{SunLing2020Cose,VietsAD2018Rtcs,CahillaneCraig2017CUfA,AcerneseF2018Coav}}.

 Some pre-processing steps are necessary.
The time series is usually down sampled from its original sampling frequency  of 16384~Hz for LIGO and 20~kHz for Virgo to a lower rate (typically 4096~Hz). Then it is band-pass filtered \textcolor{red}{because the LIGO detectors are calibrated only in the frequency band from 10~Hz to 5~kHz and the Virgo detector is calibrated in the band from 10~Hz to 8~kHz.
 In some analyses, the time series is also}
notch filtered around known instrumental noise frequencies.
Software packages for pre-processing are included in the LAL library \citep{lalsimulation} and the Gravitational Wave Open Science Center \citep{AbbottR2019Odft, Software_GWOSC}. 
Parameter estimation is often performed in the frequency domain.
After a discrete Fourier transform, 
model (\ref{modelintime}) is equivalent to the following frequency domain model:
\begin{equation}\label{modelinfrequency}
\tilde{d}_j = \tilde{d}(f_j)= \tilde{h}^{(k)}(f_j|\btheta) + \tilde{n}^{(k)}(f_j),\qquad f_j=2\pi j/T,\quad j=0,\ldots,N
\end{equation}
with
\begin{equation}\label{signalinfrequency}
\tilde{h}^{(k)}(f_j|\btheta)=\left(F^{(k)}_+(\alpha,\delta,\psi)\tilde{h}_+(f_j|\btheta) + F^{(k)}_\times(\alpha,\delta,\psi)\tilde{h}_\times(f_j|\btheta)\right) \e^{- i f_j \tau^{(k)}}
\end{equation}

\begin{figure}
\centering
\includegraphics[width=1.0\textwidth]{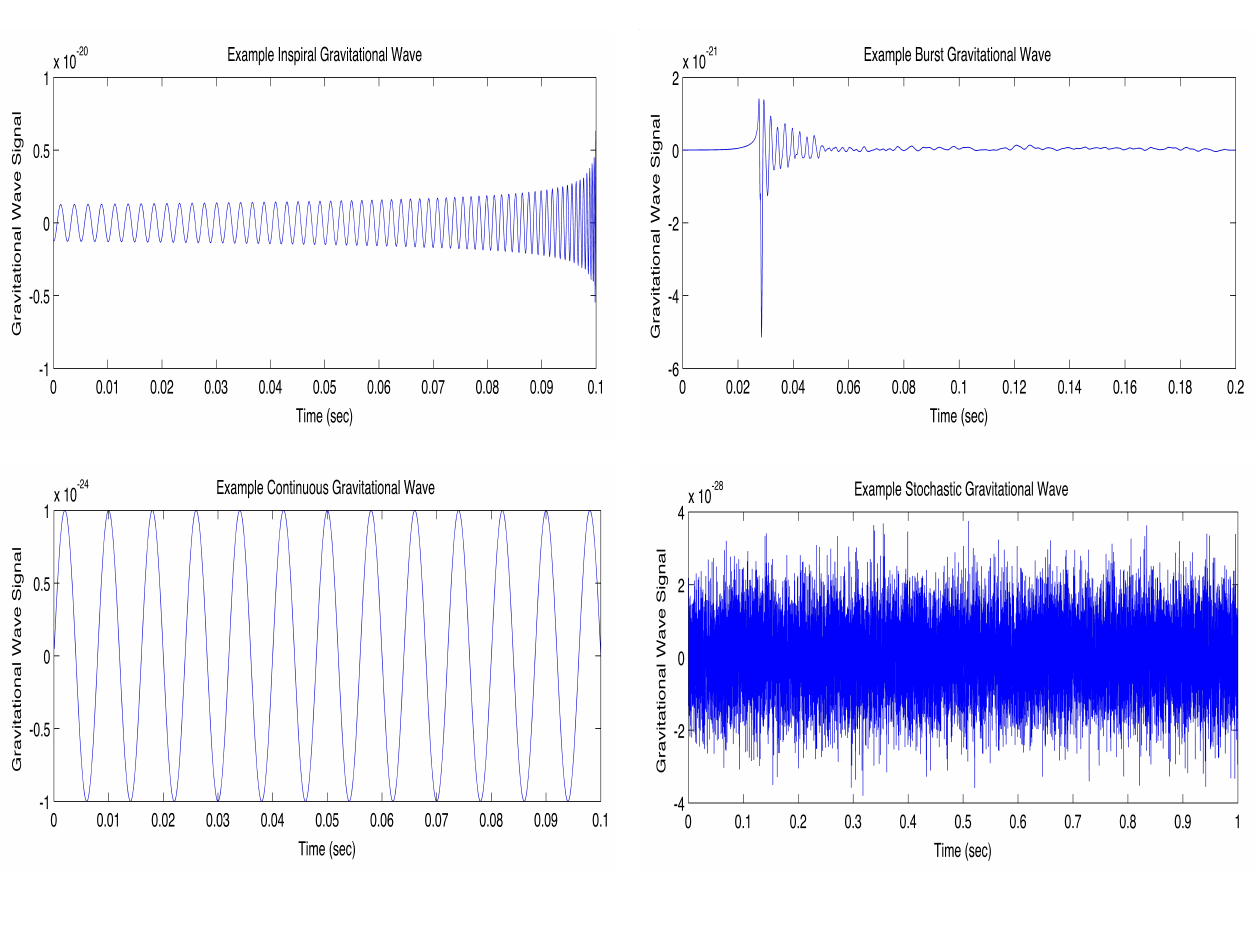}

\caption{Simulated gravitational wave signals with different waveforms: a compact binary inspiral, core-collapse supernova, continuous wave, and stochastic gravitational wave signal. Credit: A. Stuver, LIGO Scientific Collaboration \citep{4figs}.}
\label{fig:sources}
\end{figure}

The exact form of the gravitational waveform model $h_{+,\times}(t|\btheta)$
depends on the emitting source of the gravitational waves. In the following sections, we describe these waveform models from compact binary coalescences, burst signals from core-collapse supernovae, continuous signals from rapidly rotating neutron stars (pulsars), and stochastic signals from astrophysical and cosmological origins that combine to form the stochastic gravitational wave background, illustrated in Figure \ref{fig:sources}. For each of these gravitational waveform sources, we will specify the computational techniques used to estimate their parameters. In Section \ref{sec:noise} we describe methods for estimating the unknown noise spectral density in unison with the signal parameters.
\bigskip

\section{Compact Binary Coalescences}
{\label{CBC}}

To date, the Advanced LIGO and Advanced Virgo detectors have  observed
gravitational waves from the coalescence \textcolor{red}{ of dozens of compact binary systems containing black holes and also two
neutron star mergers}.  Chirp-like signals of short duration, such as 
GW150914 \citep{Abbott:2016blz}, are generated during the final stage of a  binary system as the two progenitor masses spiral in toward one another and then merge.  
Many different waveform families
for binary mergers exist in the literature. Most are parametric
waveform models obtained by solving Einstein's equations and can be
constructed within different frameworks.  They can be based on solving
the two-body dynamics in general relativity perturbatively using
post-Newtonian (PN) approximation of various orders
\citep{PhysRevD.80.084043,BlanchetLuc2014GRfP,PhysRevD.94.104015}. Two other frameworks were considered for
the analyis of GW150914: the effective-one-body (EOBNR) waveforms
\citep{BuonannoA.1999Eoat, Purrer2016,Nagar:2018zoe, OssokineSerguei2020MEWf} where higher order PN terms are
calibrated to numerical relativity (NR) simulations, and hypbrid/phenomenological (IMRPhenom)
waveforms \citep{HannamMark2014Asmo, Khan:2015jqa,PrattenGeraint2020Stcf} based on extending frequency-domain PN
expressions and hybridizing PN and EOB with NR waveforms. For the simulation of gravitational wave signals,
numerical relativity calculations have become very important~\citep{Brugmann366,doi:10.1002/andp.201800348}.
\textcolor{red}{Surrogate models of NR waveforms have been shown to be both fast and accurate \citep{VarmaVijay2019Smoh}.} 

When dealing with a binary neutron star system, or a neutron star - black hole binary,
the neutron star can be distorted from its spherical shape from the tidal gravitational
fields that it experiences. This deformation will actually increase the rate 
of the inspiral. In this case the gravitational waveform model should 
address this deformation. For example, with the examination of the 
gravitational wave data from the binary neutron star merger GW170817~\citep{TheLIGOScientific:2017qsa}
the waveform used was generated with the post-Newtonian
and EOBNR formalisms, which also included tidal deformations~\citep{PhysRevD.83.084051,AbbottB2019Potb}.

Here we exemplify the parameter
estimation using 8 s of data around the transient event
GW150914 \citep{TheLIGOScientific:2016wfe, Romero-ShawI2020Bifc} from both Advanced LIGO detectors,
comprising about ten cycles during the inspiral phase, followed by
merger and ring-down.  Merging black holes \textcolor{red}{ in a quasi-circular orbit} are described by eight {\em
  intrinsic} parameters, the masses $m_{1,2}$ and the spins $s_{1,2}$
(in terms of the dimensionless spin magnitude $a=c|s|/(Gm^2)\in [0,1]$
and orientation) of the individual black holes and an additional seven
{\em extrinsic} parameters, the luminosity distance $D_L$, the right
ascension $\alpha$ and declination $ \delta$ of the source, the
orientation in terms of the inclination angle $\iota$ between the
system's orbital angular momentum and the line of sight, and the
polarization angle $\psi$, the time $t_c$ and the phase $\phi_c$ of
coalescence. \textcolor{red}{If the spins of the masses are parallel to the orbital plane, then there can be a spin-orbit coupling that will cause a precession of the orbital plane. In this case $\iota$ is not well defined, and it is more appropriate to define the angle between the total angular momentum of the system and the line of sight, $\theta_{JN}$. We will ignore orbital precession for the remainder of this paper and continue to use $\iota$.}
 For GW150914, the orbital eccentricity of the binary system was not considered. This would have added
an additional two parameters, the magnitude and the argument of periapsis of the system.  

During the inspiral phase, the gravitational wave polarizations observed at the angle $\iota$ can be expressed at the leading order as
\begin{eqnarray}
h_+(t|\btheta)&=&A(t|\btheta)\frac{1}{2}(1+\cos^2\iota)\cos \phi(t|\btheta)\\
h_{\times}(t|\btheta)&=& A(t|\btheta)\cos\iota \sin\phi(t|\btheta)
\end{eqnarray}
where $A(t|\btheta)$ and $\phi(t|\btheta)$ are the gravitational wave amplitude and phase, respectively. The gravitational wave frequency $f$ equals twice the orbital frequency. Due to the emission of gravitational waves the binary system loses energy causing the orbital distance to decrease and the orbital frequency to increase. The phase evolution $\phi(t)$ \textcolor{red}{in the inspiral regime is well described by}  post-Newtonian theory, 
a perturbative expansion in powers of the orbital velocity $v/c$.
The first order gravitational wave frequency evolution is described by the differential equation \citep{NoAuthor2017}
\begin{equation}
\frac{df}{dt} =\frac{96}{5}\pi^{8/3}\left( \frac{G \mathcal{M}}{c^3}  \right)^{5/3}f^{11/3}
\end{equation}
where $c$ is the speed of light and $G$ Newton's gravitational
constant.  We see that the phase evolution is mainly influenced by the
chirp mass $\mathcal{M}=\frac{(m_1m_2)^{3/5}}{(m_1+m_2)^{1/5}}$ as a
result of energy loss from emitting gravitational waves. The chirp
mass can be much more accurately estimated than the individual masses.
Additional parameters such as the mass ratio $q=m_1/m_2$ (or
equivalently the symmetric mass ratio
$\eta=\frac{m_1m_2}{(m_1+m_2)^2}=\frac{q}{(1+q)^2}$) and the spin
components enter at each of the following PN orders.  This is accurate
in the inspiral phase but degrades as the black holes get closer and
eventually the full solution to Einstein's equations is needed using
numeral relativity.  Merger and ringdown depend primarily on the mass
and spin of the final black hole. Note that the observed frequency of
the signal is redshifted by a factor $(1+z)$ where $z$ is the
cosmological redshift \citep{KrolakA1987Cbot} and cannot be distinguished from a rescaling of
the masses by the same factor. Thus the source mass is obtained by
dividing the measured redshifted mass by $(1+z)$. \textcolor{red}{When the luminosity distance is estimated this can be converted to a redshift $z$ by assuming a Lambda Cold Dark Matter  ($\Lambda CDM$) cosmology and an appropriate value of the Hubble constant  \citep{Ade:2015xua}.}

After a Fourier transform of the \textcolor{red}{ strain time series}, parameter estimation
is performed in the frequency domain using the likelihood function
 (\ref{Whittlelikelihood}).  \textcolor{red}{Note that most waveform models used in practice are constructed in the frequency domain. That includes phenomenological models, surrogate models of effective-one-body waveforms and surrogate models built directly from NR waveforms.} The Bayesian model is completed
by specifying prior distributions for the parameters. For parameter
estimation of GW150914, the masses were assumed to have a uniform
prior on $[10,80]\; \textrm{M}_{\odot}$ with $m_2\leq m_1$. The spin
magnitudes $a_{1,2}$ were assumed to be uniformly distributed on
$(0,1)$ and the spin orientations isotropic on the 2-sphere. The prior
on $t_c$ was chosen to be uniform, centered at the reported time of
coalescence with a width of 0.2 s, $\phi_c$ and $\psi$ were assumed to
be uniform on $[0,2\pi]$.  As the density of sources is assumed uniform in the
cosmological co-moving volume, the prior for the source location in
the Universe was isotropic, i.e.\ the prior for $\alpha$ uniform on
$[0,2\pi]$ and the prior for  $\cos(\delta)$ uniform on $[0,1]$,
and the distance prior is uniform in Euclidian volume. 
 The prior on the cosine of the  inclination angle $\iota$ was uniform on $[0,1]$.

The product of prior and likelihood determines the posterior
distribution of the parameters using Bayes' theorem
\begin{equation}\label{eq:Bayes}
p(\btheta|\d) =\frac{ L(\d|\btheta)p(\btheta)}{\int  L(\d|\btheta)p(\btheta) d\btheta}
\end{equation}A comprehensive treatment of Bayesian inference can be found in the textbook of \cite{GelmanAndrew2014} and an introduction to Bayesian inference for astronomers in
\cite{thrane_talbot_2019}.  Evaluation of the normalizing constant in
the denominator (also called the {\em marginal likelihood} or {\em evidence}) and
calculating marginal distributions and their summary statistics of individual parameters requires
high-dimensional integration.  \textcolor{red}{To solve these high-dimensional integration problems,  computer-intensive simulation-based methods are required for several reasons:   the posterior distribution is not
tractable analytically, numerical integration is only feasible in low
dimensions, Laplace approximation \citep{GelmanAndrew2014} is suitable
only for unimodal and symmetric posteriors, and ordinary simulation
methods based on independent random draws such as importance sampling
\citep{GelmanAndrew2014}  are also only applicable effectively in low
dimensions.}
\citet{PhysRevD.58.082001} demonstrated the use of Markov chain Monte
Carlo (MCMC) methods for gravitational wave parameter estimation for posterior computation using a simple
low-order waveform model with four parameters. MCMC methods were
readily taken up by LIGO -- Virgo and increasingly sophisticated MCMC
techniques were developed to handle higher order waveform
approximations with increasing number of parameters
\citep{PhysRevD.64.022001, Pai200130,
  Christensen2004317,Rover:2006ni,Rover2007, vanderSluys:2007st,
  vanderSluys:2008qx,Veitch:2009hd, Aasi:2013jjl}.  Implemented in
LALInference is an adaptive version of the Metropolis-Hastings
algorithm \citep{Metropolis19531087} coupled with parallel tempering \citep{SwendsenRobertH.1986RMCS,Veitch:2014wba} which
is sufficiently flexible to handle any of the waveform models in the
LAL libraries. For a detailed description of the parallel tempering MH
algorithm, see \cite{vanderSluys:2007st,vanderSluys:2008qx}.

Parallel tempering uses a series of functional probability densities also known as power posterior densities, which generate a bridge between the prior and the posterior distributions.  The sampling is performed on these bridging densities, but allowing point swaps in adjacent chains, according to a certain probability.  Thus, the exploration of the posterior distribution is prevented from being stuck in certain areas of the parameter space.  This  algorithm has been implemented in software packages such as LALinference \citep{Veitch:2014wba} and BayesWave \citep{Cornish:2014kda}. Even though the samples from these intermediate bridging densities are discarded from the parameter inference process, these samples can be used to  accurately calculate the marginal likelihood \citep[e.g.,][]{PhysRevD.80.063007, Veitch:2014wba, MaturanaR:2019}.  They can also be used to build a proposal distribution based upon estimation of the kernel density and tuned to the target posterior \citep{Farr:2013tia}. The open-source Python-based parameter estimation toolkit for compact binary coalescence signals, {\tt PyCBC Inference} \citep{BiwerChristopherMichael2019PIAP}, uses the ensemble MCMC algorithm {\tt emcee} \citep{Foreman-MackeyDaniel2013:TMH} and its parallel-tempered version {\tt emcee\_pt} \citep{ptemcee}.

An alternative algorithm, also implemented in LALInference and routinely used together with the MCMC algorithm is nested sampling (NS) \citep{skilling2006}, which evaluates the evidence/marginal likelihood, i.e.\ the denominator in Bayes theorem (Eq.~\ref{eq:Bayes}). It can also be extended to generate samples from the posterior distribution at no extra cost. 
  Its particular way of exploring the parameter space allows it to work in cases in which popular MCMC methods fail \citep{Maturana:2019}.  Basically, it does this by sampling a number of points from the prior and then the one with the lowest likelihood value is replaced by a new point drawn from the prior, but restricted to have a likelihood higher than the one that is being replaced.  This procedure is repeated multiple times, allowing to keep points in different modes simultaneously and deal with complex likelihood functions.
NS was first  used  for gravitational wave searches with  ground-based observatories \citep{Veitch:Vecchio:2008a,Veitch:Vecchio:2008b}.  As it can generate samples from the posterior, it was then applied for parameter estimation and model selection for binary inspiral systems \citep{Veitch:2009hd}.
NS type algorithms have also been used for model selection and parameter estimation of space-based detectors \citep[e.g.,][]{Gair:Porter:2009, Feroz_etal:2009,Gair_etal:2010,MarsatSylvain2020EtBp}.
 NS is available in computational packages such as Bilby \citep{Ashton:2018jfp,Romero-ShawI2020Bifc} (a flexible Python-based package that also includes several MCMC samplers)  and in parallelised versions \citep{Smith:Ashton:2019}, which can be used in computing clusters.

\begin{figure}
\centering
\includegraphics[width=0.9\textwidth]{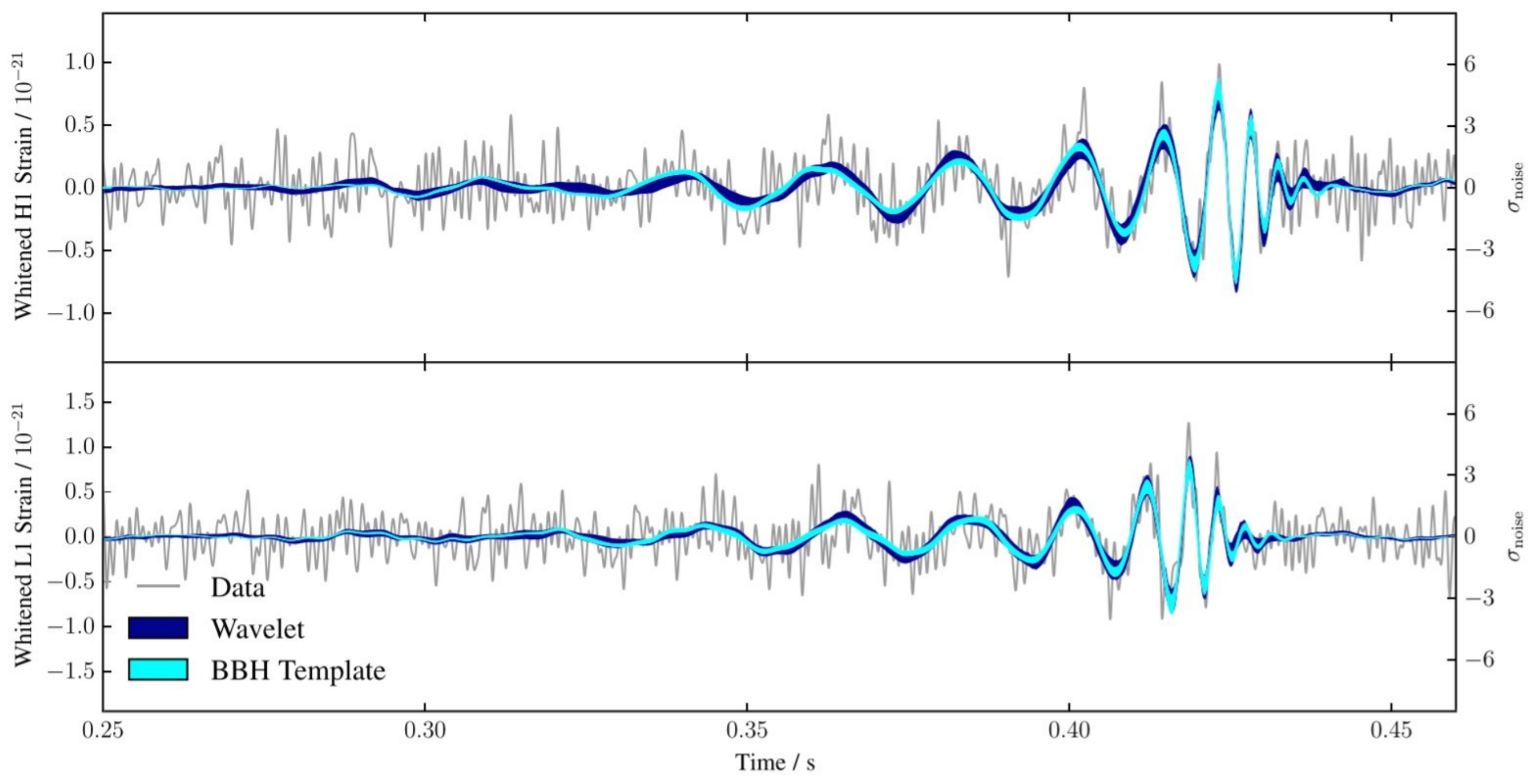}
\caption{The detector strain measurements of  GW150914~\citep{Abbott:2016blz} observed by the LIGO Hanford (H1, top) and Livingston (L1, bottom) detectors in grey. Times are shown relative to September 14, 2015 at 09:50:45 UTC. The cyan curve is the estimated signal using the IMRPhenom and EOBNR waveform templates, the blue curve a signal reconstruction based on BayesWave. Source: \cite{TheLIGOScientific:2016wfe}.}
\label{fig:GW150914}
\end{figure}

Both MCMC and NS were used to estimate the parameters of the very first gravitational wave signal GW150914 observed by Advanced LIGO, yielding consistent sets of parameter estimates \citep{TheLIGOScientific:2016wfe}. A reconstruction of the gravitational wave signal is shown in Figure \ref{fig:GW150914}. A table with parameter estimates and their standard errors can be found in Table I of \cite{TheLIGOScientific:2016wfe}. The importance that Bayesian parameter estimation played in describing the physics associated with the first direct observation of gravitational waves with \textcolor{red}{GW150914} is summarized in \citet{doi:10.1111/j.1740-9713.2016.00896.x}. 

In a compact binary coalescence event that includes at least one
neutron star -- a binary neutron star or a neutron star - black hole
merger --  electromagnetic signatures at different timescales and wavelengths are expected if the neutron star is tidally disrupted before merger. Being able to
rapidly provide an estimate of the sky location is particularly
important for multi-messenger astronomy \citep{AbbottB.P.2017Mooa}. \textcolor{red}{After the detection of GW150914, the estimate of its sky location was shared with 
63  ground- and space-based observatories covering radio, optical, near-infrared, X-ray, and gamma-ray wavelengths \citep{AbbottBp2016LABF}.}
 Fast localization of the
gravitational wave source allows a targeted follow-up by
electromagnetic telescopes as was the case for GW170817. \textcolor{red}{To this end, Bayestar has
been developed. Bayestar conditions on fixed values of the
intrinsic parameters and computes the posterior of the extrinsic parameters. This allows} an approximation of the marginal
posterior distribution of the sky location via numerical integration
\citep{PhysRevD.93.024013}. This can provide reliable sky localization and distance estimations 
within minutes after detection. Bayestar gave an initial estimate of
the position in the sky of GW170817 of 31 deg$^2$ and an estimate of
the luminosity distance of 40$\pm$ 8 Mpc \citep{GCN21513}. A more
accurate estimate of the sky location, 28 deg$^2$, was then provided by
LALInference~\citep{TheLIGOScientific:2017qsa}. \textcolor{red}{Methods for describing the three-dimensional posterior  distribution of sky location and distance have been developed \citep{SingerL.P.2016GtDM,SingerLeoP.2016Sgtd}.}

A different approach to modelling the gravitational waves of compact
binary coalescences is implemented in the BayesWave algorithm of
\cite{Cornish:2014kda}. It is not based on a physically meaningful
gravitational waveform model but aims to reconstruct the shape of any
burst signal using wavelets. 
The BayesWave reconstruction of GW150914 is displayed in Fig.~\ref{fig:GW150914}.
\cite{GhongeSudarshan2020Rgws} compare
its reconstruction properties to the reconstructions via MCMC and NS in LALInference.

Sampling algorithms such as MCMC and NS give generally accurate
parameter estimates for compact binary inspirals but can be very slow,
e.g.\ it takes days to obtain results for black hole mergers, weeks
for neutron star mergers.  Efficient but accurate approximative methods \citep{SmithRory,CanizaresPriscilla2015AGWP,PhysRevD.93.024013} can yield a significant  reduction in computation time. 
Reduced-order models (ROMs) of gravitational waveforms have reduced the computational cost of Bayesian  inference
 by  \textcolor{red}{a more efficient decomposition of the waveform using analytical insight}  \citep{PurrerMichael2015Fdro}. \cite{SmithRory} construct a ROM that includes the effects of spin precession, inspiral, merger, and ringdown in compact object binaries
utilizing the ``IMRPhenomPv2" waveform model. A fast reduced-order quadrature allows to approximate posterior distributions at greatly reduced computational costs.  A review of waveform acceleration techniques based on reduced order or surrogate models that speed up parameter estimation is provided in \cite{Setyawati_2020}.
\cite{Vinciguerra_2017} exploit the chirping behaviour of compact binary inspirals to sample sparsely for portions where the full frequency resolution is not required, \cite{ZackayBarak2018RBaF} and \cite{CornishNeil2013FFMa} use relative binning and the heterodyning principle, respectively, for fast likelihood evaluation.
Rapid parameter inference methods using grid techniques are also being developed~\citep{PhysRevD.92.023002, Lange:2018pyp}.

To speed up parameter estimation, deep
learning approaches, particularly variational autoencoders and
convolutional neural networks, have recently been explored
\citep{GeorgeDaniel2018DLfr,GabbardHunter2019Bpeu,ShenHongyu2019DaBN,ChuaAlvin2020LBpw,green2020gravitationalwave}. Deep
learning approaches train neural networks to learn the posterior
through stochastic gradient descent to optimize a loss function. The
training samples require only sampling from the prior and the
likelihood which is fast. It also has the advantage that training can
be performed offline and the estimation of parameters from an observed
gravitational wave signal becomes almost instantaneous. These methods
are still in their infancy and need to be further developed to be able
to handle the full parameter space of binary inspirals and longer
duration waveforms from multiple detectors. They hold great promise
for low-latency parameter estimation and a fast electromagnetic
follow-up.

With a whole catalogue of compact black hole mergers from the first, second, and soon third LIGO -- Virgo observations runs, it has now become feasible to infer population properties such as their merger
 rates, mass spectrum, and spin distribution, as for instance in 
 \cite{StevensonSimon2017Haog,FishbachMaya2018DtBH, WysockiDaniel2019Rpdo, ChaseKimball2020BhgI, CallisterThomas2020SaMC, AbbottB2019BBHP, smith2020inferring}.


\bigskip

\section{Unmodelled Burst Signals}
\label{sec:supernovae}

There are various potential origins of unmodelled short-duration {\it
  burst} signals with no known closed form, such as pulsar glitches,
core collapse supernovae, gamma ray burst engines, and unanticipated
sources. Amongst these, gravitational waves from core collapse
supernovae (CCSNe) are probably the most promising for observation
\citep{gossan:15}.

To date, gravitational waves from CCSNe have not been directly
observed by the network of terrestrial detectors, Advanced LIGO and
Advanced Virgo \citep{PhysRevD.101.084002}.  However, a new era in
multimessenger astronomy began with the observation of gravitational
waves from a binary neutron star inspiral (GW170817) with associated
counterparts observed across the electromagnetic spectrum
\citep{TheLIGOScientific:2017qsa}.  Much like GW170817, CCSNe are an
important source of multimessenger astronomy as they will have
associated electromagnetic, as well as neutrino counterparts~\citep{PhysRevLett.58.1490,PhysRevLett.58.1494}. The
gravitational wave signal from a CCSN will be of order 1 s or less.
LIGO and Virgo regularly search for gravitational wave signals from
CCSN~\citep{AbbottBP2019Asfs,Abbott:2019pxc}.

Like neutrinos, gravitational waves are emitted from the core of the
progenitor and carry information about the dynamics of the core
collapse and shock wave revival mechanism that leads to
explosion~\citep{Kuroda_2017}.  However, gravitational  waveforms
from CCSNe are analytically intractable due to the complex interplay
of general relativity, particle physics, and nuclear physics, meaning
template-based search methods like those employed in compact binary
coalescence pipelines are \textcolor{red}{currently} not possible.  Alternative parameter
estimation routines are needed.

The first attempt to conduct parameter estimation on CCSN gravitational wave signals
was by \citet{Summerscales_2008}, who used the maximum entropy
framework to deconvolve noisy data from multiple detectors to extract
a CCSN gravitational wave signal.  They made inferences on amplitude and phase
parameters using cross correlation between a recovered waveform and a
set of competing waveforms from the \citet{ott:2004} waveform
catalogue, where a match was defined as the model with the maximum
cross correlation to the recovered waveform.

\begin{figure}
\centering
\includegraphics[width=0.9\textwidth]{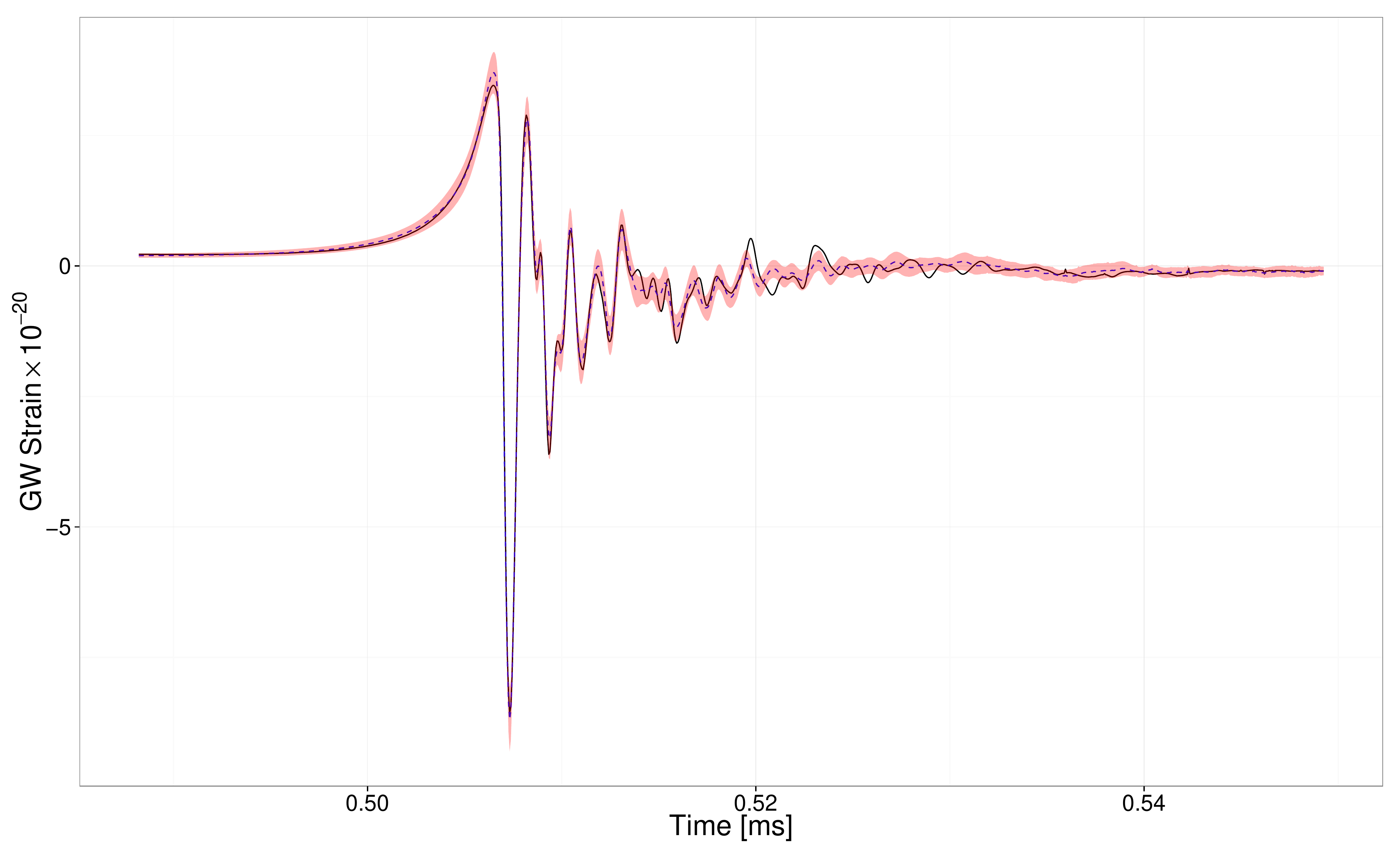}
\caption{Reconstructed CCSN signal from the \citet{dimmelmeier:2008}
  waveform catalogue using a transdimensional PCR model. The true
  signal (solid black) and model-averaged reconstruction (dashed blue)
  are overlaid with the model-averaged 90\% credible region (shaded
  pink).}
\label{fig:ccsn_recon}
\end{figure}

\citet{Heng_2009} proposed simplifying the problem using principal
component analysis (PCA) to reduce a supernova waveform catalogue
parameter space to a small number of basis vectors.  \citet{Rover2009}
extended on this by creating a Metropolis-within-Gibbs sampler to
reconstruct rotating core collapse signals using principal component
regression (PCR).  They attempted to conduct parameter estimation by
matching reconstructed signals to catalogue waveforms using a $\chi^2$
distance, but this had limited success.  \citet{EdwardsThesis}
extended the PCR Bayesian reconstruction of CCSN signals using a
birth-death reversible jump MCMC (RJMCMC) approach, allowing the
number of principal components to vary, and making use of model
averaging to handle the model selection problem.  An example of a
reconstructed CCSN waveform from the \citet{dimmelmeier:2008} waveform
catalogue can be seen in Figure~\ref{fig:ccsn_recon}.

\citet{abdikamalov:2014} used matched filtering on their newly created
waveform catalogue to infer total angular momentum from rotating CCSN
signals with errors up to 20\% for rapidly rotating progenitors and
35\% for slowly rotating cores.  They also used nested sampling to
classify precollapse differential rotation profile, with reasonable
success.

\citet{Edwards2014} demonstrated that it is possible to extract
astrophysically meaningful information encoded in the posterior
principal component coefficients in the Bayesian PCR model of
\citet{Rover2009}.  Using posterior predictive sampling, they were
able to give Bayesian credible intervals for the first time on
parameters such as the ratio of rotational kinetic energy to
gravitational potential energy of the inner core at bounce.  The
authors also used supervised machine learning methods to classify the
precollapse differential rotation profile.

\citet{Engels:2014nua} used frequentist multivariate regression and
classical hypothesis testing to analyse important astrophysical
parameters from CCSNe signals.  In contrast to the PCR approach of
\citet{Heng_2009, Rover2009} to reconstruct waveforms, the authors
used the method of least squares to find an encoded relationship
between the PC basis functions and astrophysical parameters,
identifying the most important astrophysical parameters from signals
buried in simulated detector noise.

\textcolor{red}{A recent extension of the PCA-based approach to parameter estimation of CCSN signals has been given by 
\cite{RomaVincent2019Awcs}. It includes features in the gravitational wave signal that are associated with g-modes and the standing accretion shock instability.
Rather than computing the principal components of the simulated waveform time series, it performs a PCA on the spectrograms and test the performance using simulated data for planned future detectors such as the Einstein Telescope and Cosmic Explorer.}

The supernova explosion mechanism is not fully understood.  The two
most popular and well-studied supernova explosion mechanisms are the
\textit{neutrino-driven} explosion for non-rotating and slow-rotating
progenitors and \textit{magnetorotational-driven} explosion for
rapidly rotating progenitors \citep{janka:2012}. As the gravitational
wave signals from these explosion mechanisms are morphologically
different, this has been an area of much focus for parameter
estimation studies.  This was first formulated as a classification
problem by \citet{PhysRevD.86.044023}, where they used PCR and nested
sampling \citep{skilling2006}, computing Bayesian evidence to select
the most likely explosion mechanism.  Classifying the supernova
explosion mechanism using nested sampling has been further studied by
e.g., \citet{powell:2016} using real detector noise,
\citet{powell:2017} for 3D simulations and noise transient rejection.

\cite{CoughlinM2014Mfeo}, \cite{AbbottBP2019Asfs} and \cite{10.1093/mnras/staa181} developed  Bayesian approaches for estimating the parameters of transient signals based on the time-frequency maps.
\cite{Lynch:2015yin} explored an information-theoretic approach to the
burst detection problem. Both used nested sampling as part of an
algorithm for detecting short-duration gravitational wave bursts.
The method of \citet{CoughlinM2014Mfeo} can also be used for long-duration gravitational wave
transients, 
possibly lasting up to thousands of seconds.

One of the most popular and sophisticated methods for constructing
unmodelled bursts is the BayesWave algorithm by
\citet{Cornish:2014kda}.  BayesWave uses Morlet-Gabor continuous
wavelets to construct bursts and glitches.  Under the reversible jump
MCMC framework \citep{Green1995711}, they treat the number of wavelets
as variable.  Parameter estimation routines for unmodelled bursts with
BayesWave are currently being developed, with reasonable success at
sky localisation \citep{becsy:2017}.

In line with the current trend in gravitational wave data analysis,
deep learning methods are being explored.  Similar to the deep
learning methods for glitch characterisation \citep{Zevin_2017,
  george:2018}, convolutional neural networks have started populating
the CCSN parameter estimation literature (see e.g.,
\citet{astone:2018, HengIk2019DaCo, iess:2020}) due to their success in
image classification problems.  However, deep learning approaches for
CCSNe are still in their infancy and have not developed beyond
classification problems.

\bigskip

\section{Continuous Signals}
{\label{pulsars}}

Various sources, such as binary systems far from coalescence or non--axisymmetric rotating neutron stars will emit continuous gravitational waves. 
Gravitational wave signals from rapidly rotating biaxial or triaxial neutron stars, so-called pulsars,  have been searched for but have not yet been observed by Advanced LIGO -- Advanced Virgo~\citep{AbbottB.P.2017Fsfg, AbbottB.P.2019Nsfg, AbbottBp2019Sfgw, PhysRevD.100.122002, Pisarski:2019vxw}. Pulsars  emit
gravitational waves that will likely be seen on Earth as weak
continuous  signals and are promising candidates for future observations runs. These quasi-periodic signals are of long duration with near constant amplitude and frequency. The gravitational wave signal from such an object is at twice its rotation frequency $f_s=2f_r$.
Identification of a periodic gravitational wave signal is challenging because of the weakness of the signal. But radio observations can provide information about the sky location, rotation frequency, and spindown rate of known pulsars, thus  allowing a targeted search 
in a very narrow spectral window ~\citep{Dupuis_2005}.
The observed signal,
described by 
\[h(t)= F_+(t,\psi)h_0\frac{1}{2}(1+\cos^2\iota)\,\cos\phi(t) + F_{\times}(t,\psi)h_0 \cos\iota \, \sin\phi(t),\]
depends on several unknown parameters: the overall amplitude of the gravitational wave signal  $h_0$ ,
the  polarization angle $\psi$, the angle  $\iota$  between spin axis of the pulsar and the line of sight, and the 
phase evolution $\phi(t)$. 
The response of the detector to the two polarization is given by $F_+(t,\psi)$ and $F_{\times}(t,\psi)$. \textcolor{red}{The sky position parameters $\alpha$ and $\delta$ are fixed.}
A simple slowdown model provides the rotational phase evolution of the signal via a short Taylor series expansion
\[
 \phi(t) = \phi_{0} + 2\pi \left[f_{\rm s}(T - T_{0}) + \frac{1}{2}\dot{f_{\rm s}} (T -
       T_{0})^{2}+ \frac{1}{6}\ddot{f_{\rm s}}(T - T_{0})^{3} + \ldots\right]
\]
where
$T=  t + \delta t= t + \frac{\vec{r} \cdot \vec{n}}{c}  +
\Delta{T}$ is the time of arrival of the signal at the solar system
barycenter, $\phi_{0}$ is the phase of the signal at a fiducial
time $T_{0}$, $\vec{r}$ is the position of the detector with
regard to the solar system barycenter, $\vec{n}$ is a unit vector
in the direction of the pulsar, $c$ is the speed of light, and
$\Delta{T}$ contains the relativistic corrections to the arrival
time \citep{ChristensenNelson2004Mafe}. For most pulsars, the time derivative $\dot{f_{\rm s}}$ is very small and $\ddot{f_{\rm s}}$ is often swamped by timing noise.
If $f_{\rm s}$ and $\dot{f_{\rm s}}$ are known from  radio observations, the signal can be
heterodyned by multiplying the data by $\exp[-i\phi(t)]$, low-pass filtered and resampled,
 yielding a simple model with four unknown parameters $h_0, \psi,\iota, \phi_0$. If there is an uncertainty in the frequency and frequency derivative then we
have two additional parameters, the differences between the signal and heterodyne
frequency and frequency derivatives 
 \citep{ChristensenNelson2004Mafe, Umstatter2004, Dupuis_2005}.
Early MCMC techniques for sampling from the posterior distribution used a combination of reparametrizaton, delayed rejection, and simulated annealing \cite{Umstatter2004}.
Nested sampling code for parameter estimation and model selection in targeted searches for continuous gravitational wave signals from pulsars has been developed by~\cite{PitkinM2012Ancf} and  implemented in the  {\sf LALInference} software  and  is described in detail in \cite{PitkinMatthew2017Ansc}.
\cite{AshtonGregory2018HmMf} introduce a method for the hierarchical follow-up of continuous gravitational wave candidates by leveraging MCMC optimization of the F-statistic using the affine-invariant ensemble sampler~\citep{Foreman-MackeyDaniel2013:TMH}.

\citet{PhysRevD.100.044009} highlighted that deep learning,
particularly convolutional neural networks, can be used to directly
search for continuous waves.  Though the results were much faster than
matched filtering, the approach needs to be further developed to be
more competitive (in terms of the detection probability) with existing
methods.

\bigskip

\section{Stochastic Gravitational Wave Background}
{\label{SGWB}}

The combination of all gravitational wave signals  that can not be individually resolved will make up what is called the stochastic gravitational wave background (SGWB)~\citep{Romano:2016dpx,ChristensenNelson2019Sgwb}. Analogous to the cosmic microwave background, the physical processes in the early evolution of the Universe created the {\em cosmological} SGWB. The superposition of many   unresolved signals from many independent sources such as the galactic population of white dwarf binaries, compact binary mergers,  supernovae, pulsars, magnetars, and cosmic strings  make up the {\em astrophysical} SGWB. As electromagnetic waves cannot provide information about astrophysical sources and processes any earlier than 400,000 years after the Big Bang -- the time of last scattering -- detection and estimation of the SGWB is extremely important to probe the early Universe. Unlike transient gravitational wave signals that come from a certain location in the sky, the SGWB signal will come from all directions and may or may not be isotropic and uniformly distributed across the sky~\citep{TheLIGOScientific:2016dpb,AbbottB.2019Sfti}. By and large,  the SGWB is a stochastic signal and  will  be another source of noise in a single detector,
often modelled as stationary Gaussian with mean zero and positive definite covariance matrix or spectral density to be estimated. So 
the fundamental problem is to distinguish the SGWB ``noise" from instrumental noise~\citep{AdamsMatthew2010DbaS}.
When there are several detectors such as in the network of Advanced LIGO -- Advanced Virgo, cross-correlation methods can be employed, e.g. with
observations at two detectors
\[
d^{(k)}(t)= h(t) + n^{(k)}(t), \quad k=1,2
\]
assuming independent noise components, the correlation between the observations becomes \citep{PhysRevD.46.5250}:
\[\mbox{Cov}(d^{(1)}(t), d^{(2)}(t))=\mbox{Cov}( h(t) + n^{(1)}(t), h(t) + n^{(2)}(t))=\mbox{Cov}(h(t) ,h(t)).\]
Advanced LIGO and Advanced Virgo have used these correlation methods to search for the SGWB. Even though no SGWB signal has been detected to date, upper limits have been placed on the energy density of the SGWB within the frequency range of 20 Hz to 1000 Hz~\citep{TheLIGOScientific:2016dpb,LIGOScientific:2019vic}. The energy density of the SGWB is assumed to have the form
\begin{equation}
\Omega_{GW}(f) = \Omega_{\alpha} \left(\frac{f}{f_{ref}}\right)^{\alpha} ~ .
\label{eq:Omega}
\end{equation}
 $\Omega_{GW}(f)$ is the power spectral density of the SGWB divided by the {\it closure density} of the universe $\rho_{c} = 3 H^{2}_{0}/(8 \pi G)$, and the Hubble constant is $H_{0} = 67.9$ km s$^{-1}$ Mpc$^{-1}$~\citep{Ade:2015xua}. The slope of the energy spectral density $\alpha$ is assumed to be zero for a cosmological background, and 2/3 for a background created by binary black hole and binary neutron star mergers throughout the history of the universe. A reference frequency $f_{ref}$ defines where \textcolor{red}{the amplitude}  $\Omega_{\alpha}$ is measured and reported. In the latest results reported by LIGO and Virgo based on the observations from their second observing run there was no detected SGWB, and parameter estimation methods were used to generate posterior distribution functions for $\Omega_{\alpha}$ and $\alpha$. The Bayesian parameter estimation method used by LIGO-Virgo for the SGWB search was first presented by \citet{PhysRevLett.109.171102}. For the constraint of $\alpha = 0$, a presumed cosmological background, a 95\% credible level upper limit was set at 25 Hz to be $\Omega_{0} < 6.0 \times 10^{-8}$, while for a compact binary produced background with $\alpha = 2/3$ the upper limit for $f_{ref} = 25$ Hz is $\Omega_{2/3} < 4.8 \times 10^{-8}$~\citep{LIGOScientific:2019vic}.

General relativity predicts that the polarization of gravitational waves would only have a tensor form, while alternate theories of gravity predict vector and scalar polarizations. LIGO and Virgo have searched for the presence of these alternate polarizations, and have not found a signal. Using nested sampling, a Bayes factor has been used to compare the presence of a signal to Gaussian noise. Bayes factors have been computed in searches for all three polarizations, and for the results to date for LIGO and Virgo no SGWB of any polarization is observed~\citep{Callister:2017ocg,Abbott:2018utx,LIGOScientific:2019vic}.

The future space-based interferometer space antenna LISA  will search for a SGWB in the $10^{-5}$ Hz to $10^{-1}$ Hz frequency band~\citep{2017arXiv170200786A}. LISA will yield obserations from thousands of sources of which many will remain unresolved such as gravitational waves from the galactic white dwarf binaries which will form a foreground and in addition to the instrument noise will make the estimation of the SGWB even more difficult \citep{BarackLeor2004CnfL,PhysRevD.89.022001,SachdevSurabhi2020Scbf}. 
 LISA is comprised of three coupled interferometers. A principal component-like  combination of the three output signals is made in order to eliminate the effect of this correlated noise. This is called time delay interferometry (TDI)~\citep{vinet:hal-01369369,PMID:28163627}. When expressed as the orthogonal modes ($A, E, T$), one of them, $T$, is insensitive to the gravitational wave signal at low frequencies; this channel can be used to understand the noise. The other two channels are orthogonal. This means that auto-correlation methods must be used to try to observe the SGWB~\citep{Romano:2016dpx,Smith:2019wny}, with the null channel  
employed to disentangle the detector noise from the SGWB signal. 
New search strategies and parameter estimation methods such as for instance in \cite{CapriniChiara2019Rtss} need to be developed that can be applied to  non-Gaussian, anisotropic, circularly polarized backgrounds, and backgrounds with polarization components predicted by alternative theories to general relativity~\citep{Romano:2016dpx}. This is currently an active area of research and Bayesian parametric and nonparametric methods for spectral density estimation of time series as reviewed in Section \ref{sec:noise} will be important. 
 
\bigskip

\section{Interferometer Noise}
\label{sec:noise}

As described in Section~\ref{sec:PEgeneral}, for the purpose of
estimating transient signals, the power spectral density of the noise
$n(t)$ is usually estimated `off-source', i.e.\ from a separate
stretch of data not containing the signal using Welch's method and is then
assumed to be fixed and known. For short duration signals, it might be
appropriate to assume that the noise process is stationary and
that the PSD thus does not change over time, however, for longer
duration signals, this is not a reasonable assumption. Similarly, the
assumption of {\em Gaussianity} is questionable considering the large
number of transient noise events, known as glitches, e.g.\ from
environmental sources, weather, or equipment faults.  Residuals from
parameter estimation using certain noise assumptions need to be
carefully checked as for instance described in
\cite{LIGOScientific:2019hgc}. Even if the noise is stationary, estimating the PSD from a separate data
sequence and then assuming it to be fixed for the purpose of signal
parameter estimation ignores any uncertainty in the PSD estimate and can
thus lead to biased signal parameter estimates
~\citep{ChatziioannouKaterina2019Nsem}.

For an accurate estimation of signal parameters as well as sensitive
and confident signal detection~\citep{VenumadhavTejaswi2019ANSP}, it
has been realized that a realistic modelling of the interferometer
noise is extremely important. Such a noise model should be able to
handle non-Gaussian and time-varying noise and to be included in an `on-source' method,
i.e.\ a method that estimates both signal and noise parameters
simultaneously. Various approaches have been suggested in the
literature that achieve some of these goals.

To be able to track changes in the PSD over a long period of time,
\cite{ElenaCuoco2001Opsi} suggested to split the time series into
smaller chunks and estimate the PSD using classical parametric
spectral density estimation methods based on fitting autoregressive
(AR) or autoregressive moving average models.
\cite{ZackayBarak2019DGWi} deal pragmatically with alleviating the
effect of non-stationary noise on signal detection by dividing the
matched filtering overlaps by their locally estimated standard
deviations.

For stationary Gaussian noise $\n$ with an unknown PSD, a prior model
for $S(f)$ in the Whittle likelihood (\ref{Whittlelikelihood}):
\begin{equation}\label{noiseWhittle}
L(\n)\approx\frac{1}{\det(\pi T \S)}\e^{-\frac{1}{T}\tilde{n}^*\S^{-1}\tilde{n}} = \exp\left\{-\sum_j \left[\frac{\tilde{n}(f_j)^2}{T S(f_j)} + \log(\pi T S(f_j)) \right]\right\}
\end{equation}
needs to be specified.  \cite{RoverChristian2011Mcrn} modeled the
unknown spectral density components using conjugate inverse Gamma
distributions, yielding Student-t marginal distributions for the
errors and enabling to accommodate outliers in the data. Similarly,
\cite{LittenbergTyson2013Ftco} and \cite{Veitch:2014wba} incorporate
uncertainty about the estimated PSD by an additional scale factor
$\eta_j$ for each frequency bin, i.e.\ replacing $S(f_j)$ by $\eta_j S(f_j)$, and \textcolor{red}{giving} it
a Normal prior with mean 1, where $S(f_j)$ is estimated beforehand
using the Welch method.  BayesLine, a flexible Bayesian spectral
density estimation method for Gaussian stationary noise that has been
widely applied for gravitational wave data analysis was developed by
\cite{Littenberg_2015}. BayesLine models the smooth part of the PSD by
a linear combination of cubic splines where the number as well as the
knots of the basis splines are unknown parameters.  The spectral lines
in the PSD are modeled using a sum of Lorentzians where the number,
location and line width are unknown parameters. A reversible jump MCMC
algorithm is used to sample from the posterior distribution.  This
off-source algorithm is then extended  in \cite{Cornish:2014kda} to an
on-source method, known as BayesWave, by simultaneously fitting a gravitational wave burst signal
and potential glitches, both modeled as a sum of Morlet-Gabor
continuous wavelets, see also Section \ref{sec:supernovae}. Whereas BayesWave can reconstruct the gravitational wave signal as demonstrated for instance  in 
Figure \ref{fig:GW150914} it does
not provide estimates of the physically meaningful waveform parameters. \cite{BiscoveanuSylvia2020QtEo} combined the
BayesLine model for the PSD with the physical CBC waveform models to
simultaneously estimate the signal parameters and the PSD. By
marginalizing over the PSD, the marginal posterior distributions of
the signal parameters take the full uncertainty of the PSD estimates
into account.

\textcolor{red}{ Instead of simultaneously estimating the spectral density and waveform parameters \cite{TalbotColm2020Gawa} developed a variant of Welch method by taking
the median instead of the average of periodograms over neighbouring segments. The likelihood is derived  after marginalization over the uncertainty in the median PSD estimate. The analysis is shown to be robust with respect to large outliers.}

\begin{figure}
\centering
\includegraphics[width=0.9\textwidth]{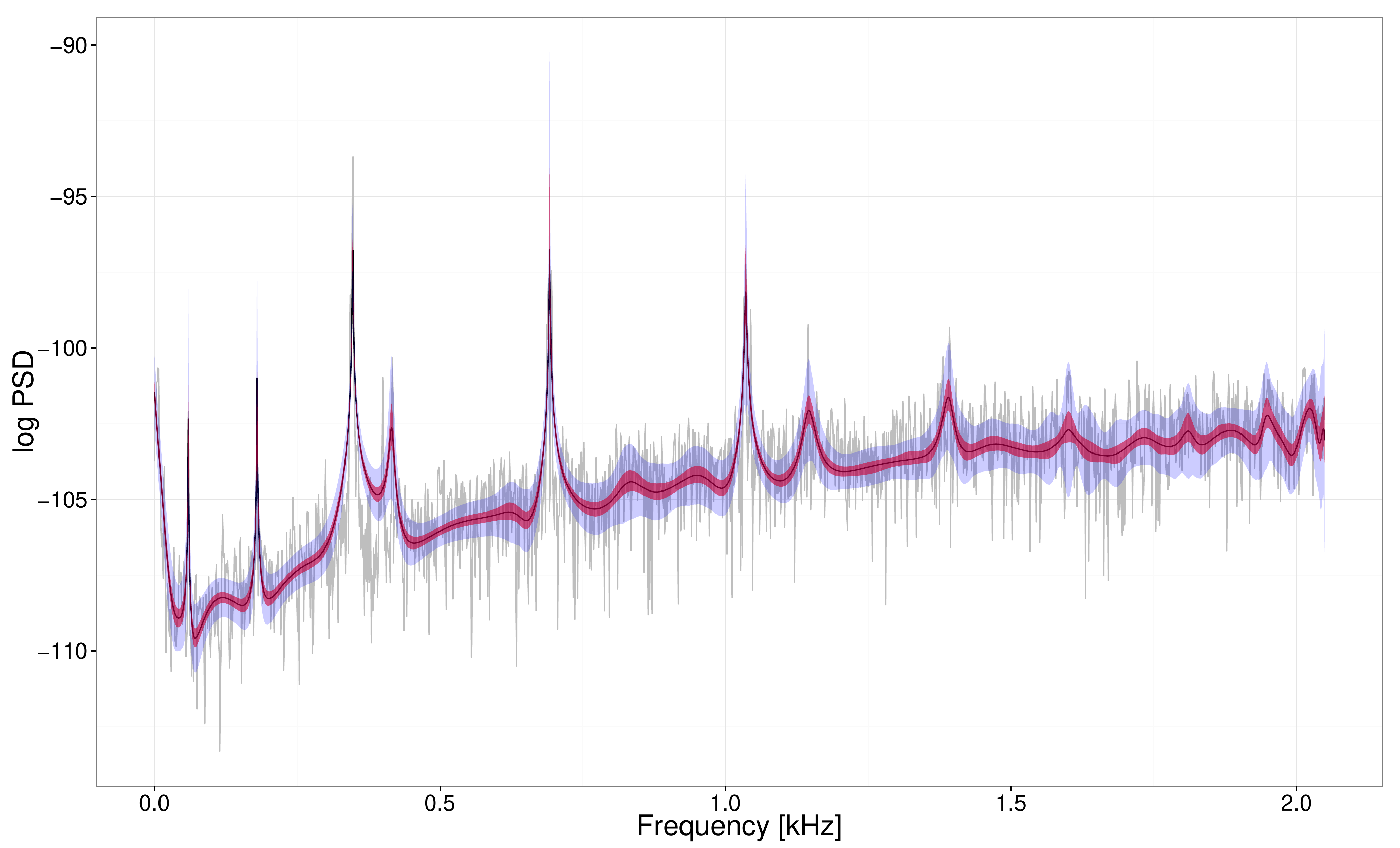}
\caption{Estimated log spectral density for a 1 s segment of Advanced LIGO S6 data.
The posterior median log spectral density estimate  using the corrected likelihood with  an AR(35) working
model (solid black), pointwise 90\% credible region (shaded red), and uniform 90\%  credible
region (shaded violet) are overlaid with the log periodogram (grey)  \citep{KirchClaudia2019BWNC}.}
\label{fig:LigoS6}
\end{figure}

Approaches based on a parametric model for the spectral density, such
as those based on fitting \textcolor{red}{autoregressive moving average} (ARMA) models, can be very efficient when the
parametric model is correctly specified but can lead to biased results
under model misspecification. Nonparametric models, on the other hand,
have much wider applicability and robustness as they do not rely on
finite-dimensional distributional assumptions.  A nonparametric
on-source approach, treating the spectral density function $S(f)$ as
an infinite-dimensional parameter and modelling this
nonparametrically using a Bernstein-Dirichlet process
prior~\citep{ChoudhuriNidhan2004BEot} has been developed by
\cite{EdwardsMatthew2015Bsps} and used to simultaneously fit rotating
core collapse supernova gravitational wave burst signals embedded in simulated
Advanced LIGO -- Advanced Virgo noise. To improve the approximation of spectral lines,
this nonparametric method was modified to use B-splines instead of
Bernstein polynomials.  By putting a Dirichlet process prior on the
knot differences, the B-spline-Dirichlet process prior was shown to be
able to accurately pick up sharp peaks and spectral lines in the data
from the LIGO S6 science run \citep{EdwardsMatthew2019Bnsd}. The method is implemented in the R
package {\tt bsplinePsd}~\citep{Edwards:bsplinePsd:2018}.  It has also been used
as on on-source model for simultaneously estimating parameters of a
non-chirping galactic white dwarf binary signal in simulated LISA data
\citep{edwards2020identifying}.  An MCMC algorithm combined with
parallel tempering was used for posterior computation. A recent
modification that reduces the computational complexity while keeping
the good approximation and coverage properties of the B-splines by
using \textcolor{red}{P-splines, i.e.\ B-splines  but with fixed knots and a smoothness penalty on the coefficients,} is given in
\cite{Maturana-RusselPatricio2019Sdeu} and implemented in the R
package {\tt psplinePsd}~\citep{psplinepackage}. By taking advantage of a well-fitting
parametric autoregressive model, \cite{KirchClaudia2019BWNC} can improve on the
Whittle likelihood approximation using a nonparametric correction of a parametric working model and prove posterior consistency. Using
the Bernstein-Dirichlet process prior for the spectral density, they
demonstrate improved performance using the same S6 LIGO noise data as
\cite{EdwardsMatthew2019Bnsd}. A spectral density estimate based on the corrected likelihood and an autoregressive working model is shown in Figure \ref{fig:LigoS6}. Sampling is based on adaptive
Metropolis-Hastings steps within the Gibbs sampler, 
implemented in the R package {\tt beyondWhittle}~\citep{Meier:beyondWhittle}. \textcolor{red}{These nonparametric approaches to spectral density estimation can be used to simultaneously  estimate the waveform parameters in a Gibbs step,  as demonstrated for instance in \cite{EdwardsMatthew2015Bsps}}.

\bigskip

\section{Conclusion}
{\label{sec:Conclusion}}

Bayesian methods for parameter estimation of gravitational wave data have proven to be essential and effective for analysing the events detected by Advanced LIGO and Advanced Virgo~\citep{LIGOScientific:2018mvr}, and will play an equally important role for the future space-based detector LISA~\citep{2017arXiv170200786A}. The two mainstays of posterior computational techniques that are routinely used and implemented in LALInference~\citep{Veitch:2014wba} are parallel tempering MCMC and nested sampling. The main problems with either algorithm when exploring the high-dimensional parameter space are potentially getting stuck in local maxima and slow mixing. These are due to the multimodality of the posterior distribution and the inherent sequential Markov chain steps of both algorithms that yield slow convergence when there are high correlations between the parameters. Any convergence acceleration methods that can yield better mixing, e.g. via adaptive MCMC methods \citep{BarberDavid2011AMcM}
and enhancements of nested sampling
\citep{FerozFarhan2013Emdw, BrewerBrendonJ2018:DNS}
will be critical. Reduced order models and surrogate waveform models \citep{SmithRory, Setyawati_2020} have and will continue to play a role in the acceleration of posterior computations.
Furthermore, the development of nonparametric approaches will allow to make the inference more robust with respect to certain assumptions. 

An alternative inferential framework which holds great promise for the future of gravitational wave analysis and has seen an  enormous increase in research activity is {\em deep learning}. This is a machine learning technique that is extremely scalable and can learn from raw data by using deep hierarchical layers 
of neural networks combined with optimization techniques based on back-propagation and gradient descent \citep{GoodfellowIan2016Dl}. 
It also allows to take account of deviations from the usual assumption of stationary Gaussian noise, as it can be trained on signals embedded in non-Gaussian and non-stationary noise.
Whereas deep learning has so far  been mainly employed for  detection and classification problems, see e.g.~ \cite{GabbardHunter2018MMFw,george:2018,Gebhard2019,Coughlin2019,Corizzo2020,Beheshtipour2020,CuocoElena2020EGSw}, recent research has a focus on the parameter estimation problem \citep{GeorgeDaniel2018DLfr,Fan2019,ChuaAlvin2020LBpw}. Its main advantage is the fact that the time-consuming training of the neural nets can be performed off-line and then potentially render the parameter estimates of an observed gravitational event in an instant.

In their third observing run, O3, Advanced LIGO and Advanced Virgo reported potential detections at a cadence of about one per week~\citep{O3_gracedb}. This rate will increase in the upcoming observing runs as the sensitivity of the detectors improves~\citep{Aasi:2013wya}. Bayesian parameter estimation will continue play a critically important role in the description of the physical systems that are producing the gravitational wave events.

\section*{Funding Information}
RM gratefully acknowledges support by the James Cook Fellowship from Government funding, administered by the Royal Society Te  Ap\={a}rangi and  DFG Grant KI 1443/3-2.
The work of NC was supported by US NSF grant PHY-1806990.


\section*{Acknowledgments}

{\label{749861}}
We thank Claudia Kirch for fruitful discussions regarding nonparametric approaches to power spectral density estimation. We thank Sharan Banagiri, Christopher Berry, Tito Dal Canton, Michael P\"urrer, and John Veitch for comments on the manuscript. This paper has been assigned LIGO document P2000177.





\bibliographystyle{apacite}
\bibliography{bibliography}

\end{document}